\documentclass[twocolumn,american,english]{revtex4-2}
\usepackage[T1]{fontenc}
\usepackage[latin9]{luainputenc}
\usepackage{amsmath}
\usepackage{amssymb}
\usepackage{babel}
\begin{document}
\title{Projected Augmented Waves (PAW) motivated mixed basis sets for small
molecules}
\author{Garry Goldstein}
\address{garrygoldsteinwinnipeg@gmail.com}
\begin{abstract}
The success behind many pseudopotential methods, such as the Projected
Augmented Waves (PAW) and the Phillips-Kleinman pseudopotential methods,
is that these methods are nearly all electron methods in disguise.
For the Phillips-Kleinman and PAW pseudopotential methods we show
that there is an explicit all electron reformulation (which is nearly
equivalent). In the all electron reformulation, as part of the basis
set, there are regular low wavevector basis wave functions (plane
waves) and several, specially chosen, high wavevector basis wave functions
that are specialized to the atomic environment relevant to nuclei
of the substances studied. Using this as motivation, here we propose
a new, PAW method motivated, basis set for small molecules, where
we use the LO or lo (Localized Orbitals) basis wave functions, and
pair them with a Gaussian basis, e.g. Gaussian Type Orbitals (GTO),
for a hybrid basis for small molecules. Contracted Gaussian functions
are also possible (CGFs). Several different, but related, LO (lo)
basis wave functions are considered. We also show how to extend this
idea to Slater basis set, Slater type orbitals (STO), instead of GTO
orbitals, combined with LO or lo orbitals. This is done using shape
functions.
\end{abstract}
\maketitle

\section{\protect\label{sec:Introduction}Introduction}

It is of paramount importance to make further progress on the basis
set problem for Density Functional Theory (DFT) for small molecules.
Small molecules larger then diatomic: carbon dioxide, water, ammonia,
methane, ethyne, ethene, ethane to name a few have a variety of uses
in industry sciences and engineering \citep{Blinder_2019,Sjostedt_2000}.
While it is true that there has been as much progress in optimizing
basis sets for small molecule DFT calculations, similarly to the basis
sets that have paved breakthroughs in solid state DFT \citep{Martin_2020,Marx_2009,Singh_1991,Andersen_1975,Andersen_1984,Andersen_2003,Louks_1967,Korringa_1947,Khon_1954,Michalicek_2013,Michalicek_2014,Singh_2006,Sjostedt_2000,Skriver_1984,Soler_1989,Soler_1990}
- for crystalline solids - further progress is needed to reach chemical
accuracy \citep{Koch_2001}. Much of the progress has been made using
ideas based on Linear Combinations of Atomic Orbitals (LCAO) \citep{Levine_2014,Blinder_2019,Helgaker_2000},
where a good basis for the entire molecule consists of linear combinations
of basis functions for the individual atoms. There are many candidate
LCAO basis sets:
\begin{itemize}
\item Gaussian basis sets (GTO) based on cartesian Gaussian Type Orbitals
(GTO) which are of the form of Gaussians multiplied by polynomials
in the cartesian co-ordinates \citep{Helgakar_1995,Helgaker_2000,Feller_1990,Blinder_2019,Hehre_1972,Hariharan_1973}.
This is advantageous as four center Coulomb integrals are efficiently
done for this basis set (see Appendix \ref{sec:Gaussian-basis-sets}).
\item Slater basis sets where Slater Type Orbitals (STOs) are used in the
LCAO method. These are spherical harmonics multiplied by exponentially
decaying radial wave functions. It is possible to have multiple wave
functions per angular momentum channel - so called multi-zeta basis
sets. STO's have advantages over GTOs as they have the correct exponentially
decaying behavior at infinity and the right Kato cusp at the origin
\citep{Helgaker_2000,Blinder_2019}, however four center integrals
are much harder - usually leading to shape function methods \citep{Koch_2001}.
\item Contracted basis sets (Contracted Gaussian Functions (CGF)) where
multiple Gaussians are combined into one basis element used to mimic
STO basis elements while retaining much of the ability to efficiently
study four centre integrals \citep{Helgakar_1995,Helgaker_2000,Feller_1990,Blinder_2019}.
\item Numerical basis sets where one solves the Kohn Sham (KS) problem numerically
and uses these wave functions as basis wave functions.
\end{itemize}
Furthermore plane waves, outside the LCAO concept, are often used
to study periodically arranged arrays of molecules, with achievable
distances between the molecules on the order of 10 $nm$, so that
interactions between the molecules may be reasonably well neglected
as a first approximation leading to a study of nearly single molecules.
The most common electronic structure method for this situation (plane
wave basis sets for crystalline arrays of molecules) is the Projected
Augmented Waves (PAW) \citep{Blochl_1994,kresse_1996,Martin_2020,Marx_2009}
method. 

Here we focus on the LCAO method. Motivated by our reformulations
of pseudopotential methods in DFT (see Section \ref{sec:Many-pseudopotentials-are})
we propose a new (LCAO) hybrid basis set. Pseudopotentials, at their
core, map a problem into another easier problem. The exact all electron
wavefunction for the system is mapped onto a smooth (pseudized) wavefunction
that may well be represented by a small number of plane waves (is
softer - easier), that is it is not the all electron wavefunction
but simplified - smoothed out. Mappings are then made between the
pseudized solution and the all electron wavefunction leading to a
solution of the KS problem. Two of the pre-eminent examples of pseudopotentials
are the Phillips-Kleinman and PAW pseudopotentials. Here we show that
these pseudopotential methods are nearly all electron methods in disguise,
where the all electron problem basis set includes smooth plane wave
wave functions (with a low cutoff) and specialized wave functions
for orbitals near atomic nuclei. In this all electron formulation
the two methods, PAW and Phillips-Kleinman, are nearly equivalent.
This however is not the main thrust of this work, but it is the main
motivation.

Motivated by the observations described in the previous paragraph
and detailed in Section \ref{sec:Many-pseudopotentials-are}, and
the extensive numerical success of the PAW method \citep{Martin_2020,Marx_2009},
we propose new basis sets for small molecules, where localized LO
or lo \citep{Goldstein_2024,Singh_1991,Singh_2006,Sjostedt_2000}
basis wave functions are augmented with Gaussian wave functions -
GTOs. GTOs are chosen because Gaussians have been extensively used
for the problem of LCAO as various configuration integrals, including
four center ones, which are well known for them \citep{Helgaker_2000,Blinder_2019,Martin_2020}
thereby simplifying the calculation of the Hamiltonian matrix and
overlap used in the KS problem - in particular of the four center
problem for Coulomb interactions part of the Hamiltonian which is
often a bottleneck for many calculations \citep{Helgaker_2000}. Furthermore
we show how to combine LO and lo basis elements with STO's for enhanced
accuracy (but with shape functions for the coulomb problem). These
constructions are done in a manner similar to Phillips-Kleinman and
PAW all electron basis sets described in the paragraph above (see
also Section \ref{sec:Many-pseudopotentials-are}). Furthermore multiple
types of LO (lo) basis wave functions are possible. Thereby we propose
that this new hybrid basis set is efficient for DFT and Hartree-Fock
(HF) calculations of small molecules.

\section{\protect\label{sec:Many-pseudopotentials-are}Many pseudopotential
methods are nearly all electron methods in disguise (motivation)}

In this section we reformulate PAW and the Phillips-Kleinman methods
as all electron methods - nearly exactly. We show the basis set of
these near equivalent all electron methods is composed of plane waves
and special wave functions that are chosen to represent states near
the core (near the nuclei). We show the two methods can be equivalent
in their all electron reformulation - depending on the exact choice
of the localized wave functions near the nuclei (cores) for both methods
and cutoffs.

\subsection{\protect\label{sec:Phillips-Kleinman-as}Phillips-Kleinman is nearly
an all electron method in disguise}

We now recall the Orthogonalized Plane Wave (OPW) wave functions used
for the Phillips-Kleinman method. We know that \citep{Singh_2006,Martin_2020}:
\begin{align}
 & \left|\phi_{OPW}\left(\mathbf{k}+\mathbf{K}\right)\right\rangle \nonumber \\
 & =\left|\frac{1}{\sqrt{V}}\exp\left(i\left(\mathbf{k}+\mathbf{K}\right)\cdot\mathbf{r}\right)\right\rangle \nonumber \\
 & -\sum_{i\mu}\left\langle \phi_{i\mu}\left(\mathbf{r}\right)\mid\frac{1}{\sqrt{V}}\exp\left(i\left(\mathbf{k}+\mathbf{K}\right)\cdot\mathbf{r}\right)\right\rangle \left|\phi_{i\mu}\left(\mathbf{r}\right)\right\rangle \label{eq:OPW-1}
\end{align}
Here $\left|\phi_{i\mu}\left(\mathbf{r}\right)\right\rangle $ are
localized wave functions relevant to the single site problem and $\mu$
labels the atoms. Here $V$ is the volume of the primitive lattice
cell and $\mathbf{k}$ is a wavevector in the first Brillouin zone
while $\mathbf{K}$ is a reciprocal lattice vector. Now the main claim
of the Phillips-Kleinman method (greatly numerically supported \citep{Singh_2006,Martin_2020})
is that for a reasonable number of plane waves the exact, all electron,
KS wavefunction $\left|\psi_{n}\right\rangle $ can be written as
\begin{equation}
\left|\psi_{n}\right\rangle \cong\sum_{\mathbf{K}}c_{\mathbf{k}}^{\mathbf{K}}\left|\phi_{OPW}\left(\mathbf{k}+\mathbf{K}\right)\right\rangle \label{eq:OPW}
\end{equation}
for some reasonable cutoffs $\mathbf{K}_{max}$ that means that 
\begin{align}
\left|\psi_{n}\right\rangle  & \in Span\left\{ \left|\phi_{OPW}\left(\mathbf{k}+\mathbf{K}\right)\right\rangle \right\} \nonumber \\
 & \subset Span\left\{ \left\{ \left|\frac{1}{\sqrt{V}}\exp\left(i\left(\mathbf{k}+\mathbf{K}\right)\cdot\mathbf{r}\right)\right\rangle \right\} ,\left\{ \left|\phi_{i\mu}\left(\mathbf{r}\right)\right\rangle \right\} \right\} \label{eq:Spna_OPW}
\end{align}
As such in some situations its better and in many ways simpler to
do an all electron calculation with the basis being given by: 
\begin{equation}
\left\{ \left\{ \left|\frac{1}{\sqrt{V}}\exp\left(i\left(\mathbf{k}+\mathbf{K}\right)\cdot\mathbf{r}\right)\right\rangle \right\} ,\left\{ \left|\phi_{i\mu}\left(\mathbf{r}\right)\right\rangle \right\} \right\} \label{eq:Basis-2}
\end{equation}
As such Phillips-Kleinman is nearly an all electron method in disguise,
this will motivate our basis set constructions in Section \ref{sec:Alternative-to-AGPAW}.
Below we shall see below that PAW is very similar. 

\subsection{\protect\label{subsec:PAW}PAW is nearly an all electron method in
disguise}

\subsubsection{\protect\label{sec:PAW-review}PAW review}

In the PAW formulation of the KS problem, we study the highly oscillatory
eigenwavefunction of the KS problem near the nucleus of an atom using
plane waves. In order to obtain a low cutoff (which can be as low
as \textasciitilde 60 Rydberg in energy for some PAW applications),
and as such a tolerable basis set, we must have that the wavefunction
we represent using plane waves (the pseudized wavefunction) is very
smooth. The key idea of PAW is then to introduce:
\begin{equation}
\left|\psi_{n}\right\rangle =\hat{\mathcal{T}}\left|\tilde{\psi}_{n}\right\rangle \label{eq:Transform}
\end{equation}
where $\hat{\mathcal{T}}$ is some linear transformation (specified
below) and $\left|\psi_{n}\right\rangle $ is the exact KS wave functions
(solutions of the KS equations) while $\left|\tilde{\psi}_{n}\right\rangle $
are the pseudized wave functions we can approximate efficiently using
plane waves at low cutoffs. Now we construct a set of augmentation
spheres and we wish for $\hat{\mathcal{T}}$ to be the identity outside
the augmentation spheres (centered around nuclei) , so we write: 
\begin{equation}
\hat{\mathcal{T}}=\mathbb{I}+\sum_{\mu}S_{\mathbf{\mu}}\label{eq:Transform-1}
\end{equation}
Here $\mu$ are the atomic augmentation spheres. Now we want the pseudized
wave functions $\left|\tilde{\psi}_{n}\right\rangle $ to be smooth
inside the sphere and we know that the exact KS wavefunction can be
reasonably efficiently approximated by a single atom, atomic wavefunction
$\left|\phi_{i\mu}\right\rangle $ near the atomic nuclei. As such
if we choose $S_{\mathbf{\mu}}$ to be 
\begin{equation}
S_{\mu}\left|\tilde{\psi}_{n}\right\rangle =\sum c_{i\mu}^{n}\left|\phi_{i\mu}\right\rangle -\sum c_{i\mu}^{n}\left|\tilde{\phi}_{i\mu}\right\rangle \label{eq:Operator}
\end{equation}
where $\left|\tilde{\phi}_{i\mu}\right\rangle $ are smooth wave functions.
Where we choose 
\begin{equation}
c_{i\mu}^{n}=\left\langle \tilde{p}_{i\mu}\mid\tilde{\psi}_{n}\right\rangle \label{eq:c_i,n}
\end{equation}
Here $\left|\tilde{p}_{i\mu}\right\rangle $ are the projector wave
functions. As such we have that: 
\begin{equation}
\left|\psi_{n}\right\rangle =\hat{\mathcal{T}}\left|\tilde{\psi}_{n}\right\rangle =\left|\tilde{\psi}_{n}\right\rangle +\sum_{i}\left(\left|\phi_{i\mu}\right\rangle -\left|\tilde{\phi}_{i\mu}\right\rangle \right)\left\langle \tilde{p}_{i\mu}\mid\tilde{\psi}_{n}\right\rangle \label{eq:Basis}
\end{equation}
Furthermore the pseudized wave functions $\left|\tilde{\psi}_{n}\right\rangle $
may be well represented by plane waves with low cutoff.

\subsubsection{\protect\label{sec:Reformulation-of-PAW}Near exact reformulation
of PAW as an all electron method}

As reviewed in Section \ref{sec:PAW-review} the main claim of PAW
(which has been extensively numerically verified) is that there is
a good expansion in plane waves for the PAW wavefunction given by:
\begin{equation}
\left|\psi_{n}\right\rangle \cong\left|\tilde{\psi}_{n}\right\rangle +\sum_{i}\left(\left|\phi_{i\mu}\right\rangle -\left|\tilde{\phi}_{i\mu}\right\rangle \right)\left\langle \tilde{p}_{i\mu}\mid\tilde{\psi}_{n}\right\rangle .\label{eq:Wavefunctions}
\end{equation}
That is the wavefunction $\left|\tilde{\psi}_{n}\right\rangle $ is
made of a reasonably small number of plane waves of the form $\frac{1}{\sqrt{V}}\exp\left(i\left(\mathbf{k}+\mathbf{K}\right)\cdot\mathbf{r}\right)$
in particular we have that: 
\begin{equation}
\left|\psi_{n}\right\rangle \in Span\left\{ \left\{ \frac{1}{\sqrt{V}}\exp\left(i\left(\mathbf{k}+\mathbf{K}\right)\cdot\mathbf{r}\right)\right\} ,\left\{ \left|\phi_{i\mu}\right\rangle \right\} ,\left\{ \left|\tilde{\phi}_{i\mu}\right\rangle \right\} \right\} \label{eq:Span}
\end{equation}
to significant accuracy, for small number of plane waves. Now because
$\left|\tilde{\phi}_{i}\right\rangle $ is very smooth we have that
\begin{equation}
\left|\tilde{\phi}_{i\mu}\right\rangle \in Span\left\{ \frac{1}{\sqrt{V}}\exp\left(i\left(\mathbf{k}+\mathbf{K}\right)\cdot\mathbf{r}\right)\right\} \label{eq:Span_smooth}
\end{equation}
As such the main claim of PAW can be rewritten as 
\begin{equation}
\left|\psi_{n}\right\rangle \in Span\left\{ \left\{ \frac{1}{\sqrt{V}}\exp\left(i\left(\mathbf{k}+\mathbf{K}\right)\cdot\mathbf{r}\right)\right\} ,\left\{ \left|\phi_{i\mu}\left(\mathbf{r}\right)\right\rangle \right\} \right\} \label{eq:Smaller_span}
\end{equation}
for a reasonable number of plane waves. Since the basis set of PAW
and Phillips-Kleinman when formulated as all electron methods are
similar, both methods should yield, in their all electron formulation,
similar accuracy and require basis sets of the same size. They are
equivalent within the all electron formulation when they use the same
localized orbitals $\left|\phi_{i\mu}\left(\mathbf{r}\right)\right\rangle $
and the same cutoff.

\section{\protect\label{sec:Alternative-to-AGPAW}Gaussian LO (LO), Slater
LO (SLO) and Plane Wave LO (PLO) basis sets}

Here we would like to work backwards, we wish to formulate an all
electron method as the inverse of a pseudopotential method where we
use specialized wave functions (with high wave numbers) near the nuclei
and many regular, smooth, wave functions for the whole space. We focus
on small molecules and crystalline solids. Based on the great success
of PAW we can argue that these new basis sets increase the accuracy
of current LCAO basis sets for molecules with limited additional computational
costs and furthermore they do not rely on crystalline arrays of molecules
where intermolecular interactions can not be completely eliminated.
Below we will simply describe the final answer - that is relevant
basis sets.

\subsection{\protect\label{subsec:Glo}Glo and GLO for small molecules}

Here G in Glo and GLO stands for Gaussians while lo and LO for localized
orbitals and the two types of elements (Gaussians and LO (lo) basis
wave functions) are combined into one basis set. We consider the lo
basis set \citep{Goldstein_2024}:
\begin{equation}
\Phi_{lo}^{\mu Elm}=u_{l}^{E_{l}}\left(\left|\mathbf{r}-\mathbf{r}_{\mu}\right|\right)\Theta\left(S_{l}^{\mu}-\left|\mathbf{r-r}_{\mu}\right|\right)Y_{lm}\left(\widehat{\mathbf{r}-\mathbf{r}_{\mu}}\right)\label{eq:Glo}
\end{equation}
where $S_{l}^{\mu}$ is chosen to make the wavefunction continuous
everywhere \citep{Goldstein_2024}. Here: 
\begin{equation}
\left[-\frac{d^{2}}{dr^{2}}+\frac{l\left(l+1\right)}{r^{2}}+\bar{V}_{KS}\left(r\right)\right]ru_{l\mu}^{E_{l}}\left(r\right)=E_{l}ru_{l\mu}^{E_{l}}\left(r\right)\label{eq:Schrodinger_equation}
\end{equation}
and $\bar{V}_{KS}\left(r\right)$ is the spherically average KS potential
or Hartree-Fock (HF) single particle potential. Furthermore $E_{l}$
is chosen at the binding energy of the single atom system (other options
are possible for example one could use the eigen-energies from the
previous iteration to self consistency in the KS problem \citep{Goldstein_2024(3),Goldstein_2024(4)}
for the energies $E_{l}$). Alternatively we can use LO basis sets
\citep{Goldstein_2024}:
\begin{align}
\Phi_{LO}^{\mu Elm} & =u_{l}^{E_{l}}\left(\left|\mathbf{r}-\mathbf{r}_{\mu}\right|\right)\Theta\left(S_{l}^{\mu}-\left|\mathbf{r-r}_{\mu}\right|\right)Y_{lm}\left(\widehat{\mathbf{r}-\mathbf{r}_{\mu}}\right)\nonumber \\
 & +B_{l}^{\mu}\dot{u}_{l}^{E_{l}}\left(\left|\mathbf{r}-\mathbf{r}_{\mu}\right|\right)\Theta\left(S_{l}^{\mu}-\left|\mathbf{r-r}_{\mu}\right|\right)Y_{lm}\left(\widehat{\mathbf{r}-\mathbf{r}_{\mu}}\right)\label{eq:GLO}
\end{align}
where $B_{l}^{\mu}$ and $S_{l}^{\mu}$ are chosen to make the wavefunction
continuous and continuously differentiable everywhere \citep{Goldstein_2024}.
Here $\dot{u}_{l}^{E_{l}}=\frac{\partial}{\partial E}u_{l}^{E_{l}}$.
Higher derivative terms are also possible \citep{Goldstein_2024},
also regular LO and lo is possible \citep{Singh_1991,Singh_2006,Sjostedt_2000}
as well as atom Density Functional Theory (DFT) based wave functions
- that is solutions to the radially averaged KS problem for a single
atom \citep{Martin_2020}. Based on our experience with PAW a very
small number of wave functions such as in Eq. (\ref{eq:GLO}) will
suffice. We also add a large number of Gaussians to our basis set:
\begin{equation}
G_{l,m,\sigma}^{\mu}=\frac{1}{\mathcal{N}_{l}}\exp\left(-\frac{\left|\mathbf{r-r}_{\mu}\right|^{2}}{2\sigma}\right)\left|\mathbf{r-r}_{\mu}\right|^{l}Y_{lm}\left(\widehat{\mathbf{r}-\mathbf{r}_{\mu}}\right)\label{eq:Gaussian}
\end{equation}
to the molecular basis set. The wave functions $G_{l,m,\sigma}^{\mu}$,
are smooth and replace the plane waves we have seen before in all
electron formulations of pseudopotential methods. Here $\mathcal{N}_{l}$
is a normalization constant and cartesian Gaussians are also possible
\citep{Helgakar_1995,Helgaker_2000}. This is a basis set for many
small molecules. Furthermore Gaussians Coulomb integrals may be efficiently
computed, see \citep{Boys_1950,Helgaker_2000} and Appendix \ref{sec:Gaussian-basis-sets}
so there are ``few'' Coulomb integrals involving complicated functions
\citep{Goldstein_2024(2)}. Furthermore Eqs. (\ref{eq:Product_formula})
may be used, in the case when there is only one difficult to handle
LO (lo) wavefunction, on two of the remaining Gaussians thereby simplifying
that case to a three center problem. Overall if we consider the number
of LO or lo orbitals as order one (highly reasonable for small molecules
based on or experience with PAW \citep{Blochl_1994,kresse_1996,Martin_2020})
this becomes a three center problem at worst. We note that contracted
Gaussian Functions (CGF) are also possible \citep{Koch_2001,Helgaker_2000}.

\subsection{\protect\label{subsec:Slo-and-SLO}Slo and SLO basis sets}

We now consider a second hybrid basis set made of Slater type orbitals
and localized orbitals. We first consider basis functions of the form
given in Eqs. (\ref{eq:Glo}) or (\ref{eq:GLO}). We augment them
with basis wave functions of the form: 
\begin{equation}
S_{l,m,\xi}^{\mu}=\frac{1}{\mathcal{N}_{l}}\exp\left(-\xi\left|\mathbf{r-r}_{\mu}\right|\right)\left|\mathbf{r-r}_{\mu}\right|^{l}Y_{lm}\left(\widehat{\mathbf{r}-\mathbf{r}_{\mu}}\right)\label{eq:Gaussian-1}
\end{equation}
While four center integrals are very difficult with this basis set
three center integrals are numerically manageable \citep{Koch_2001}.
Below we describe practical methods to achieve accurate three center
decompositions into shape functions.

\subsubsection{\protect\label{subsec:Practical-methods-for}Practical methods for
implementation (three center decompositions)}

We consider standard shape functions $\omega_{\kappa}$ \citep{Koch_2001}
for the density and add to them the following additional shape functions:
\begin{equation}
\omega_{\mu Elm}^{E'l'm'}\left(\mathbf{r}\right)=\Phi_{lo/LO}^{\mu E'l'm'*}\left(\mathbf{r}\right)\cdot\Phi_{lo/LO}^{\mu Elm}\left(\mathbf{r}\right)\label{eq:Densities}
\end{equation}
this allows for hybridization between the atomic orbitals - such as
for example the $sp^{3}$ hybridization common to Carbon (if this
effect is not important for the molecule just the diagonal terms suffice).
Furthermore for notational reasons we absorb the label $\mu;Elm;E'l'm'$
into $\kappa$. We introduce the projectors $P_{\lambda}$ such that
\begin{equation}
\int d^{3}\mathbf{r}P_{\lambda}^{*}\left(\mathbf{r}\right)\omega_{\kappa}\left(\mathbf{r}\right)=\delta_{\lambda\kappa}\label{eq:Delta}
\end{equation}
Where for example we may choose: 
\begin{align}
M_{\alpha\beta} & =\int d^{3}\mathbf{r}\omega_{\alpha}^{*}\left(\mathbf{r}\right)\omega_{\beta}\left(\mathbf{r}\right)\nonumber \\
P_{\lambda}\left(\mathbf{r}\right) & =\sum_{\kappa}\left[M^{-1}\right]_{\lambda\kappa}\omega_{\kappa}\left(\mathbf{r}\right)\label{eq:Overlaps}
\end{align}
Then if we introduce the single particle density $\rho\left(\mathbf{r}\right)$
we may write that \citep{Koch_2001}
\begin{align}
\rho\left(\mathbf{r}\right) & \cong\sum_{\kappa}\left[\int d^{3}\mathbf{r}'\rho\left(\mathbf{r}'\right)P_{\kappa}^{*}\left(\mathbf{r}'\right)\right]\omega_{\kappa}\left(\mathbf{r}\right)\equiv\sum_{\kappa}c_{\kappa}\omega_{\kappa}\left(\mathbf{r}\right)\nonumber \\
c_{\kappa} & =\int d^{3}\mathbf{r}'\rho\left(\mathbf{r}'\right)P_{\kappa}\left(\mathbf{r}'\right)\label{eq:Density_decomposition}
\end{align}
We now introduce more notation for the single particle basis set where
$\left\{ \nu\right\} =\left\{ \mu,Elm\right\} \cup\left\{ \mu,l,m,\xi\right\} $
and write the eigenstates are given by 
\begin{equation}
\left|\lambda_{n}\right\rangle =\sum_{\nu}\lambda_{n\nu}\left|\psi_{\nu}\right\rangle \label{eq:Basis-1}
\end{equation}
We may now introduce \citep{Koch_2001}: 
\begin{equation}
P_{\mu\nu}=\sum_{n}f\left(\varepsilon_{n}\right)\lambda_{n\mu}^{*}\lambda_{n\nu}\label{eq:Four_point}
\end{equation}
Then the density is given by: 
\begin{align}
\rho\left(\mathbf{r}\right) & =\sum_{\mu\nu}P_{\mu\nu}\psi_{\mu}^{*}\left(\mathbf{r}\right)\psi_{\nu}\left(\mathbf{r}\right)\nonumber \\
\rho\left(\mathbf{r}\right) & \cong\sum_{\kappa}\sum_{\mu\nu}P_{\mu\nu}\left[\int d^{3}\mathbf{r}'\psi_{\mu}^{*}\left(\mathbf{r}'\right)\psi_{\nu}\left(\mathbf{r}'\right)P_{\kappa}^{*}\left(\mathbf{r}'\right)\right]\omega_{\kappa}\left(\mathbf{r}\right)\label{eq:Expansion}
\end{align}
Which is a three center integral. Furthermore the KS equation is now
given by: 
\begin{align}
H_{\alpha\beta}^{KS} & =\left\langle \psi_{\alpha}\mid\frac{-\nabla^{2}}{2m}-\sum_{\mu}\frac{Z_{\mu}e^{2}}{\left|\mathbf{R}_{\mu}-\mathbf{r}\right|}+V_{XC}\left(\mathbf{r}\right)\mid\psi_{\beta}\right\rangle \nonumber \\
 & +\sum_{\kappa}c_{\kappa}\int d^{3}\mathbf{r}_{1}\int d^{3}\mathbf{r}_{2}\omega_{\kappa}\left(\mathbf{r}_{1}\right)\frac{e^{2}}{\left|\mathbf{r}_{1}-\mathbf{r}_{2}\right|}\psi_{\alpha}^{*}\left(\mathbf{r}_{2}\right)\psi_{\beta}\left(\mathbf{r}_{2}\right)\label{eq:Three_center}
\end{align}
Which is another three center integral. As such it is possible to
do practical calculations for small molecules within DFT for SLO and
Slo basis sets.

\subsection{\protect\label{subsec:Plo-and-PLO}Plo and PLO basis sets for crystalline
solids}

Here P in Plo and PLO stands for plane wave while lo and LO stands
for localized orbitals. We first consider basis functions of the form
given in Eqs. (\ref{eq:Glo}) or (\ref{eq:GLO}) . We also add plane
wave 
\begin{equation}
\chi_{\mathbf{k}+\mathbf{K}}=\frac{1}{\sqrt{V}}\exp\left(i\left(\mathbf{k}+\mathbf{K}\right)\cdot\mathbf{r}\right)\label{eq:Plane_waves}
\end{equation}
to the solid state basis set. This is a basis set for many crystalline
solids. However there are many competitor basis sets \citep{Andersen_1975,Andersen_1984,Andersen_2003,Khon_1954,Korringa_1947,Louks_1967,Martin_2020,Michalicek_2013,Michalicek_2014,Singh_1991,Singh_2006,Sjostedt_2000,Skriver_1984,Soler_1989,Soler_1990}.

\section{\protect\label{sec:Conclusions}Conclusions}

In this work - motivated by the fact that many pseudopotential methods
are nearly all electron methods in disguise with basis sets of the
form of slowly oscillating basis wave functions (plane waves) and
special high momentum basis wave functions to account for the environment
of the nucleus - we proposed several related new basis sets for small
molecules and crystalline solids. In this basis set we use the LO
or lo wave functions introduced recently in \citep{Goldstein_2024}
(although regular LO or lo basis wave functions will also do quite
well \citep{Singh_1991,Singh_2006,Sjostedt_2000}) combined either
with Gaussians (GTOs and CGFs) or Slater basis sets - STO - (for molecules)
or plane waves (for solids) to obtain a total basis for the system.
This would allow for initial calculations for many molecules as the
four center Coulomb integral calculation is done for Gaussians and
two center integrals have been done for very generic functions \citep{Blinder_2019,Goldstein_2024(2),Helgaker_2000}
and for Slater type orbitals it is possible to introduce appropriate
density shape functions (which we extended). This should help further
open small molecules for theoretical DFT or HF like calculations and
explorations.

\appendix

\section{\protect\label{sec:Gaussian-basis-sets}Gaussian basis sets (review)}

\selectlanguage{american}%
We are interested in four center integrals for the Coulomb problem
\citep{Helgakar_1995,Helgaker_2000}: 
\begin{align}
 & I_{\alpha\beta\gamma\delta}=\nonumber \\
 & \int d^{3}\mathbf{r}_{1}\int d^{3}\mathbf{r}_{2}\frac{1}{\left|\mathbf{r}_{1}-\mathbf{r}_{2}\right|}\varphi_{\alpha}^{*}\left(\mathbf{r}_{1}\right)\varphi_{\beta}\left(\mathbf{r}_{1}\right)\varphi_{\gamma}^{*}\left(\mathbf{r}_{2}\right)\varphi_{\delta}\left(\mathbf{r}_{2}\right)\label{eq:Four_center}
\end{align}
This problem scales as $\sim N^{4}$ (where $N$ is the size of the
basis set) and must be evaluated for non-crystalline materials (molecules)
where the Weinert method does not directly apply \citep{Singh_2006}.
As such it is often the bottleneck for many calculations \citep{Helgakar_1995,Helgaker_2000}.
We can introduce spherical Gaussians to simplify these integrals:
\begin{align}
 & \varphi_{\alpha/\beta/\gamma/\delta}=\nonumber \\
 & =G_{lm,n,\mathbf{R},p}\left(\mathbf{r}\right)\nonumber \\
 & =Y_{lm}\left(\widehat{\mathbf{r}-\mathbf{R}}\right)\exp\left(-p\left(\mathbf{r}-\mathbf{R}\right)^{2}\right)\left|\mathbf{r}-\mathbf{R}\right|^{2n+l}\label{eq:Spherical_Gaussians}
\end{align}
We can also introduce cartesian Gaussians: 
\begin{align}
 & G_{a,b,c,\mathbf{R},p}\left(\mathbf{r}\right)\nonumber \\
 & =\exp\left(-p\left(\mathbf{r}-\mathbf{R}\right)^{2}\right)\left(x-\mathbf{R}_{x}\right)^{a}\left(y-\mathbf{R}_{y}\right)^{b}\left(z-\mathbf{R}_{z}\right)^{c}\label{eq:Cartesian_Gaussians}
\end{align}
We now have that \citep{Helgaker_2000}:
\begin{equation}
G_{lm,n,\mathbf{R},p}\left(\mathbf{r}\right)=\sum_{a+b+c=l+2n}A_{abc}^{lmn}G_{a,b,c,\mathbf{R},p}\left(\mathbf{r}\right),\label{eq:Transform-2}
\end{equation}
for some constants $A_{abc}^{lmn}$ \citep{Helgaker_2000}, so we
might as well focus on cartesian Gaussians for the wave functions
in the integral in Eq. (\ref{eq:Four_center}). Now we have that \citep{Helgaker_2000}:
\begin{align}
 & G_{a,b,c,\mathbf{R},p}\left(\mathbf{r}\right)G_{a',b',c',\mathbf{R}',p'}\left(\mathbf{r}\right)\nonumber \\
 & =\sum_{ABC}P_{ABC}^{abc,\mathbf{R},\mathbf{R}'}G_{ABC\bar{\mathbf{R}},p+p'}\left(\mathbf{r}\right)\label{eq:Product_formula}
\end{align}
for some constants $P_{ABC}^{abc,\mathbf{R},\mathbf{R}'}$ \citep{Helgaker_2000},
where 
\begin{equation}
\bar{\mathbf{R}}=\frac{p\mathbf{R}+p'\mathbf{R}'}{p+p'}\label{eq:Center_mass}
\end{equation}
As such for cartesian Gaussians, up to linear sums, we may as well
focus on two center integrals of the form:
\begin{align}
 & I_{\alpha\beta\gamma\delta}=\nonumber \\
 & =\int d^{3}\mathbf{r}_{1}\int d^{3}\mathbf{r}_{2}\frac{1}{\left|\mathbf{r}_{1}-\mathbf{r}_{2}\right|}G_{abc,\mathbf{R},p}\left(\mathbf{r}_{1}\right)G_{a'b'c',\mathbf{R}',p'}\left(\mathbf{r}_{2}\right)\nonumber \\
 & =\sum\kappa_{abc}^{ABC}\frac{\partial^{A+B+C}}{\partial\mathbf{R}_{x}^{A}\partial\mathbf{R}_{y}^{B}\partial\mathbf{R}_{z}^{C}}\kappa_{a'b'c'}^{A'B'C'}\frac{\partial^{A'+B'+C'}}{\partial\mathbf{R}'{}_{x}^{A'}\partial\mathbf{R}'{}_{y}^{B'}\partial\mathbf{R}'{}_{z}^{C'}}\times\nonumber \\
 & \times\int d^{3}\mathbf{r}_{1}\int d^{3}\mathbf{r}_{2}\frac{1}{\left|\mathbf{r}_{1}-\mathbf{r}_{2}\right|}G_{000,\mathbf{R},p}\left(\mathbf{r}_{1}\right)G_{000,\mathbf{R}',p'}\left(\mathbf{r}_{2}\right)\label{eq:Integral}
\end{align}
We now introduce the Boys' function \citep{Boys_1950,Helgaker_2000}:
\begin{equation}
F_{0}\left(x\right)=\int_{0}^{1}\exp\left(-xt^{2}\right)dt\label{eq:Boys_function}
\end{equation}
Then we have that: 
\begin{align}
 & \int d^{3}\mathbf{r}_{1}\int d^{3}\mathbf{r}_{2}\frac{1}{\left|\mathbf{r}_{1}-\mathbf{r}_{2}\right|}G_{000,\mathbf{R},p}\left(\mathbf{r}_{1}\right)G_{000,\mathbf{R}',p'}\left(\mathbf{r}_{2}\right)\nonumber \\
 & =\sqrt{\frac{4\alpha}{\pi}}F_{0}\left(\alpha\left|\mathbf{R}-\mathbf{R}'\right|^{2}\right)\label{eq:Boys_integral}
\end{align}
With: 
\begin{equation}
\alpha=\frac{p\cdot p'}{p+p'}\label{eq:Alpha}
\end{equation}
As such four center integrals may be efficiently evaluated with Gaussian
wave functions.

\selectlanguage{english}%
.

\end{document}